\documentclass[prb,oneocolumn,showpacs,showkeys,amsmath,amssymb]{revtex4-1}
\usepackage{refstyle}
\usepackage{graphicx}
\usepackage{dcolumn}
\usepackage{bm}
\usepackage{subfigure}
\usepackage{bm}
\usepackage{tabularx}
\usepackage{lettrine}
\usepackage{type1cm}
\usepackage{amsmath}
\usepackage{amsthm}
\DeclareMathOperator{\sech}{sech}

\begin{document}
\title{Multiband Superconductivity in Lu$_3$Os$_4$Ge$_{13}$}
\author{Om Prakash}
\email[Om Prakash]{op1111shukla@gmail.com}
\address{Department of Condensed Matter Physics and Materials Science,
Tata Institute of Fundamental Research, Mumbai-400005, India}
\author{A. Thamizhavel, S. Ramakrishnan}
\address{Department of Condensed Matter Physics and Materials Science,
Tata Institute of Fundamental Research, Mumbai-400005, India}

\begin{abstract}

Intermetallic R$_3$T$_4$X$_{13}$ series consists of cage like structure and have been in 
focus due to their unconventional electronic ground states. In this work, we report the 
normal and superconducting state properties of a high quality single crystal of Lu$_3$Os$_4$Ge$_{13}$. 
Lu$_3$Os$_4$Ge$_{13}$ belongs to the above mentioned series and crystallizes in a cubic 
crystal structure with the space group $\it{Pm\bar{3}n}$. Using electrical transport, magnetization 
and heat capacity measurements, we show that Lu$_3$Os$_4$Ge$_{13}$ is a type-II multi-band 
superconductor ($T{_c} =3.1$~K) with unusual superconducting properties. The analysis of 
the low temperature heat capacity data suggests that Lu$_3$Os$_4$Ge$_{13}$ is a moderately 
coupled multi-band BCS superconductor with two gaps ($2\Delta / {k{_B}T{_c}} = 3.68 \pm {0.04} ~\& ~0.34 \pm {0.02}$) in the superconducting state. The 
dc-magnetization ($M-H$) shows a large reversible region in the superconducting state similar to the vortex liquid phase observed in high-$T{_c}$ 
superconductors. The value of the Ginzburg number $G_{i}$ suggests that the thermal fluctuations, 
though small as compared to those in high-$T{_c}$ cuprates, may play an important role in 
the unpinning of the vortices in this compound. The electronic band structure calculations show 
that three bands cross the Fermi level and constitute a complex Fermi surface in Lu$_3$Os$_4$Ge$_{13}$.
\end{abstract}


\maketitle
\section{Introduction}

Ternary intermetallic compounds with the formula R$_3$T$_4$X$_{13}$, where R is a rare-earth 
element, T transition metal and X semi-metallic/semiconducting element, have significantly contributed 
to the understanding of the physics of strongly correlated materials.  
These compounds show a variety of unusual magnetic ground states and superconductivity at low 
temperatures \cite{Maple2014}. One such structure with no metalloids was reported by Remeika 
{\it et al} \cite{Remeika1980}. These compounds (R$_3$T$_4$X$_{13}$) 
crystallize in a cubic structure (space group $\it{Pm\bar{3}n}$) and 
feature cage-like environment within the unit cell. Amongst these, La, Yb, and Th based compounds are 
superconducting, whereas, Gd and Eu based 
compounds show magnetic transition. By replacing Sn by Ge and Rh by Ru in R$_3$Rh$_4$Sn$_{13}$, 
Segre and Braun \cite{Braun1981} reported superconductivity and magnetic ordering. Our studies 
on polycrystalline samples of R$_3$Ru$_4$Ge$_{13}$ series
(R=Y, Ce, Pr, Nd, Ho, Er, Dy, Yb, Lu) showed \cite{Ramakrishnan1996,Ramakrishnan1995,Ramakrishnan1993} that 
the series exhibit unusual physical properties. These properties could 
be considered as those belonging to the semi-metals or low band-gap semiconductors. 
Studies on Lu and Ru based polycrystalline samples \cite{Ramakrishnan1993} reported superconductivity 
below 2.3~K and 2.8~K respectively. These compounds are also
reported as good thermo-electric materials \cite{Kong2007}. The multi-valley nature of 
the band structure \cite{Cohen1964}, cage like coordination in the crystal structure and 
occurrence of superconductivity make iso-structural Lu$_3$Os$_4$Ge$_{13}$ an interesting 
compound to look for unusual superconductivity. Bulk studies show that Lu$_3$Os$_4$Ge$_{13}$ 
is a multi-band superconductor and it shows a large reversible region in the magnetization 
(M-H) data, which has been usually observed in high $T{_c}$ superconductors \cite{Blatter1994}.
In this report, we present detailed study of the anomalous superconducting properties of 
Lu$_3$Os$_4$Ge$_{13}$ single crystal, supported by the electronic band structure calculations.

\section{Methods}
\subsection{Sample preparation and characterization}

\begin{figure}[h]
\includegraphics[width=16cm,angle=0]{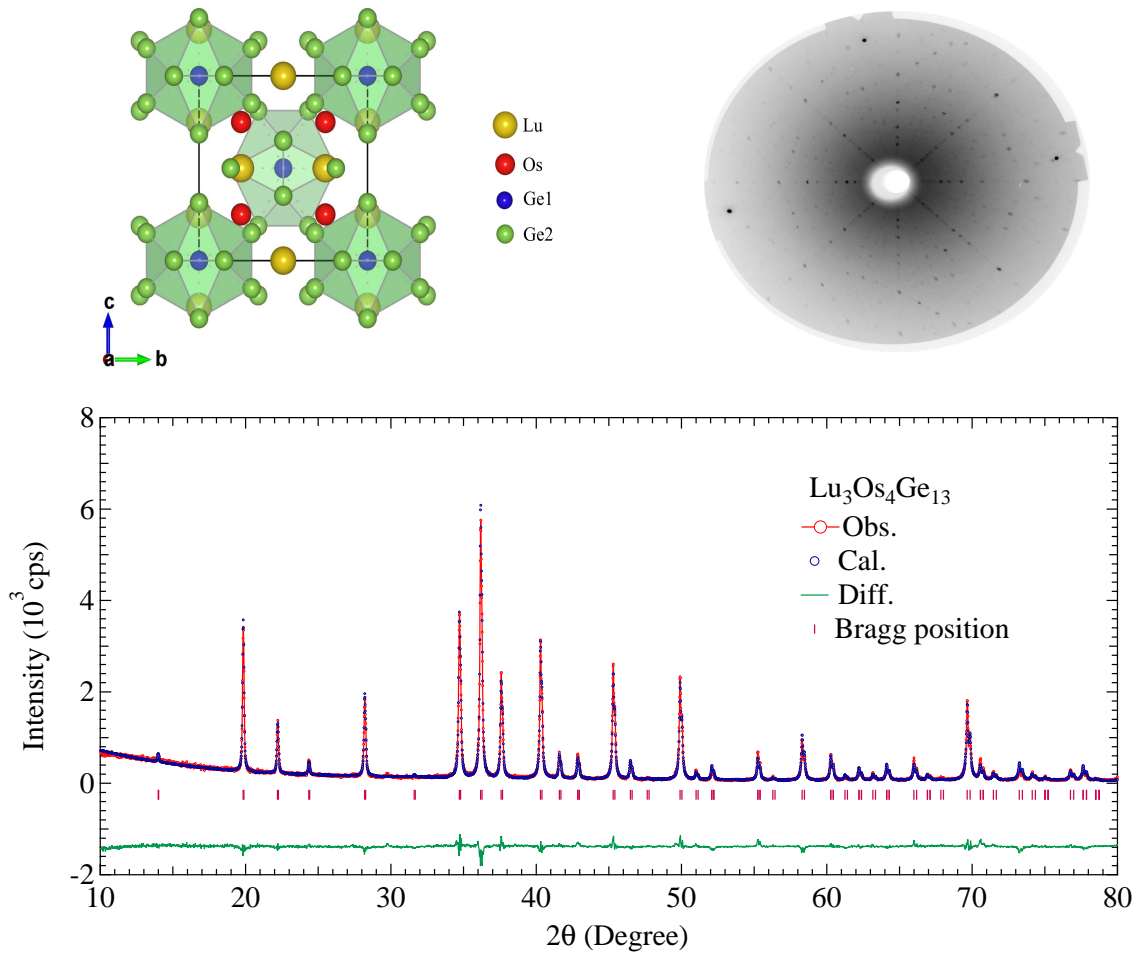}
\caption{(Color online) 1. Top left panel: Crystal structure of Lu$_3$Os$_4$Ge$_{13}$ projected along (100) plane. 2. Top right panel: Laue diffraction pattern for (100) plane of Lu$_3$Os$_4$Ge$_{13}$. 3. Bottom panel: Rietveld fit of the powder XRD data of Lu$_3$Os$_4$Ge$_{13}$. The unit cell consists of cage like crystal structure as shown in the left panel. The Rietveld analysis and circular spots in the Laue diffraction confirm the high quality, single phase and single-crystalline nature of the grown crystal.}
\label{fig:fig1}
\end{figure}

The single crystal of Lu$_3$Os$_4$Ge$_{13}$ was grown using the Czochralski crystal pulling method in a 
tetra-arc furnace under inert Argon atmosphere. Stoichiometric mixture (10~g) of highly pure elements 
(Lu: 99.99\%, Os: 99.99\%, Ge: 99.99\%) was melted 4-5 times in the same furnace to make a 
homogeneous polycrystalline sample. The single crystal was pulled from the polycrystalline melt 
using a tungsten seed rod at 
the rate of $10$~mm/h for about $6$~hrs to get $5-6$~cm long cylindrical shaped crystals with 
$3-4$~mm diameter. Lu$_3$Os$_4$Ge$_{13}$ crystallizes in cubic crystal structure with space group $\it{Pm\bar{3}n}$
(space group number 223) with 40 atoms per unit cell (2 formula unit). The structure is similar 
to that of iso-structural compound Y$_3$Ru$_4$Ge$_{13}$ \cite{Prakash2013}. The unit cell contains two inequivalent Ge sites.
One can visualize Lu$_3$Os$_4$Ge$_{13}$ structure as a unit 
consisting of three substructures: edge-sharing Ge1(Ge2)12 icosahedra, Lu-centered cubotahedra R(Ge2)12, and 
corner-sharing Os(Ge2)6 trigonal prisms. The corner-sharing Os(Ge2)6 trigonal prisms create ``cages" containing a 
Ge1 atom similar to the cages observed in Skutterudites\cite{Shi2007}. The crystal structure of Lu$_3$Os$_4$Ge$_{13}$ is shown in Fig.~\ref{fig:fig1}. The room temperature powder XRD data were analyzed by structural Rietveld refinement \cite{Carvajal1993} 
using the Fullprof program  (shown in Fig.~\ref{fig:fig1}) and the refinement confirmed single phase nature of the sample. 
The global ${\chi{^2}}=2.74$ is obtained from the refinement. The values of lattice constants 
obtained from the refinement are $a=b=c=8.94585$ ($\pm$ 0.00022 $\AA$).
\begin{table}[h]
\caption{\label{tab:XRD} Crystal structure parameters obtained from the Rietveld refinement of the room 
temperature powder x-ray diffraction data of Lu$_3$Os$_4$Ge$_{13}$. Profile reliability factor $R_{p} = 17.5\% $, 
weighted profile $R$-factor $R_{wp}=17.2\%$, Bragg R-factor $= 8.46\%$ and R$_{f}$-factor $= 7.44\%$ were obtained from the best fit.}

\begin{ruledtabular}
\begin{tabularx}{\textwidth}{cccccc}
\multicolumn{2}{l}{Structure} &\multicolumn{3}{l} {Cubic} \\
\multicolumn{2}{l}{Space group} & \multicolumn{2}{l} {$\it{Pm\bar{3}n}$ (No. 223)}\\
\multicolumn{2}{l}{Lattice  parameters} \\
\hline
\multicolumn{2}{l}{\hspace{0.8cm} $a$ ({\AA})}                                  &  8.94585(22)  \\        
\multicolumn{2}{l}{\hspace{0.8cm} $V_{\rm cell}$  ({\AA}$^{3}$)}        & 715.921( 0.030)  \\
\\
\multicolumn{2}{l}{Atomic coordinates} \\
\hline
Atom & Wyckoff & \hspace{-0.5 cm} $x$ & \hspace{-0.7 cm} $y$ & \hspace{-0.5 cm} $z$\\
&  position \\
Lu   &  6c & \hspace{-0.5 cm} 0.25 & \hspace{-0.7 cm} 0 & \hspace{-0.5 cm} 0.50 \\
Os   &  8e & \hspace{-0.5 cm} 0.25 &  \hspace{-0.7 cm} 0.25 & \hspace{-0.5 cm} 0.25 \\
Ge1  &  24k & \hspace{-0.5 cm} 0 & \hspace{-0.7 cm} 0.31206(30) & \hspace{-0.5 cm} 0.14965(32)\\
Ge2  &  2a & \hspace{-0.5 cm} 0 & \hspace{-0.7 cm} 0 & \hspace{-0.5 cm} 0 \\
\end{tabularx}
\end{ruledtabular}
\end{table}
The crystal was characterized using various experimental techniques such as room temperature powder 
X-ray diffraction (PXRD), electron probe micro-analyzer (EPMA) and energy dispersive 
X-ray spectroscopy (EDX) and confirmed to be a well defined single crystal with no trace of impurity phases. 
The EPMA and EDX characterizations were done on the well polished surfaces and confirmed the proper 
stoichiometry ($3:4:13$) and single phase nature of Lu$_3$Os$_4$Ge$_{13}$ compound. The single crystals 
were oriented along the crystallographic direction [100] using Laue back reflection method using the 
Huber Laue diffractometer and cut to the desired shape and dimensions using a spark erosion cutting machine.

\subsection{Measurement techniques}

The electrical resistivity was measured using standard four-probe technique in a home made 
setup. 40~$\mu$m diameter Au wires were used for making electrical connections using indium solder. The contact resistance is of the order of 10m$\Omega$. The zero 
magnetic field data was recorded in the temperature range 1.6-300~K using LR700 resistance bridge. 
The electrical resistivity was measured in constant magnetic fields (0-5.5~T) in a Cambridge Magnetic Refrigeration (CMR) mFridge refrigrator setup from 
0.1-4.2~K for determining the upper critical field $\mu{_0}H{_{c2}}(T)$. The CMR setup can reach to base temperature of $\approx$ 100mK using adiabatic demagnetization of paramagnetic salt pills. Magnetic susceptibility was measured 
using a commercial superconducting quantum interferometer device (SQUID) magnetometer (MPMS5, Quantum Design, 
USA) in a constant magnetic field of 10~mT; the sample was cooled down to 1.8~K in zero-magnetic field 
and then the magnetic field was applied, followed by warming to 5~K (this is called the zero field 
cooled "ZFC" data). Then the sample was cooled down to 1.8~K in the magnetic field of 10~mT to take 
field cooled (FC) data. The heat capacity was measured using Physical Property Measurement System (PPMS), 
equipped with He${^3}$ dilution refrigerator in the temperature range 0.05-4~K and 1.8-300K by a time-relaxation 
method in different magnetic fields from 0-7~T. 

\subsection{Band structure calculations}

The electronic band structure calculations were performed by density functional theory (DFT) using WIEN2k 
code with a full-potential linearised augmented plane-wave and local orbitals (FP-LAPW + lo) basis, 
\cite{Blaha2001, Madsen2001} together with Perdew-Burke-Ernzerhof (PBE) parametrization \cite{Perdew1996} 
of the generalized gradient approximation (GCA), with no spin-orbit coupling. 
The plane wave cutoff parameter $R{_{MT}}K{_{MAX}} = 7$ was taken with 5000 k-points (the choice of number 
of k points varies with symmetry of crystal structure and lower symmetry structure may require larger 
number of k-point sampling). The program xCrysden was used for calculating and plotting bands and Fermi surface.

\section{Results and Discussion}

\begin{figure}[h]
\includegraphics[width=16cm,angle=0]{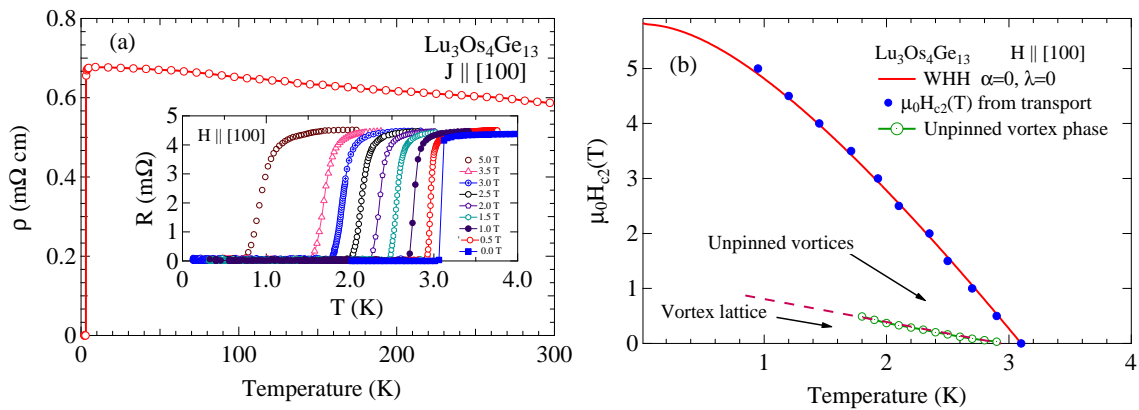}
\caption{(Color online) (a) Temperature dependence of the electrical resistivity (${\rho}$) for 
the current parallel to the [100] direction of Lu$_3$Os$_4$Ge$_{13}$ from 2-300~K. The inset 
shows the effect of the magnetic field on the superconducting transition temperature. 
(b) Temperature dependence of the upper critical $\mu{_0}H{_{c2}}$. 
The red curve correspond to the simulated WHH expression (~(\ref{eq:6})) in the dirty limit 
for $\alpha =0$, $\lambda =0$. 
The open green circles shows the phase boundary between the vortex lattice and unpinned vortex phase, 
as observed in the magnetization data.}
\label{fig:fig2}
\end{figure}

Fig.~\ref{fig:fig2}(a) shows the electrical resistivity of Lu$_3$Os$_4$Ge$_{13}$ for 
current parallel to the [100]-direction 
measured in the temperature range from 2-300~K. A negative temperature coefficient of resistivity
$(\frac{d\rho}{dT}<0)$ is observed in the normal state. This can result from several 
mechanisms such as site disorder, electronic correlations 
and low energy phonon scattering. The compound under study has a cage like structure with 
a loosely bound Ge atom within the cage, which gives rise to the low energy soft phonon modes 
resulting in significant electron-phonon scattering at low temperatures. The resistivity becomes zero at the onset of 
superconductivity below 3.1~K. Inset in Fig.~\ref{fig:fig2}(a) 
shows the change in the superconducting transition temperature in different magnetic fields. 
The width of the superconducting transition ($\Delta T$) increases with increasing magnetic field. 
The transition temperature is taken at the point where the resistivity drops to 50$\%$ of the 
normal state value. Fig.~\ref{fig:fig2}(b) shows the temperature dependence of the upper 
critical field $\mu{_0}Hc{_2}$.  A linear relationship is observed between $H{_{c2}}$ and 
temperature in the proximity of the transition temperature ($T{_c}$ at H = 0).

In type-II superconductors, an external magnetic field leads to the Cooper pair breaking 
via two mechanisms, orbital and spin-paramagnetic effects \cite{Helfand1966, Chandrasekhar1962, Clogston1962}
(the latter also known as Pauli paramagnetic limiting effect). The orbital pair breaking is 
related to the emergence of Abrikosov vortices. The orbital limiting field refers 
to the magnetic field at which the vortex cores fill the whole volume and is given by the formula,
\begin{equation}\label{eq:1}
{H{_{c2}}^{\mathrm{orb}}(0)}=\Phi{_0}/{2\pi }{\xi{^2}},
\end{equation}
where $\xi$ is the Ginzburg-Landau coherence length and $\Phi{_0}$
 ($ = {hc}/{2e} = 2.07\times{10^{-15}}$Tm${^2}$), 
is the magnetic flux quantum. For the single band BCS superconductors, ${H{_{c2}}^{\mathrm{orb}}(0)}$ 
is derived from the slope of the $H{_{c2}}$(T)-T phase boundary at $T{_c}$, given by,
\begin{equation}\label{eq:2}
\mu{_0}Hc{_2}^{\mathrm{orb}}(0)=-0.69T{_c}\frac{dHc{_2}}{dT}\vert{_{T=T{_c}}},
\end{equation}
in the dirty limit and 
\begin{equation}\label{eq:3}
\mu{_0}Hc{_2}^{\mathrm{orb}}(0)=-0.73T{_c}\frac{dHc{_2}}{dT}\vert{_{T=T{_c}}},
\end{equation}
in the clean limit of the Werthamer-Helfand-Hohenberg (WHH) theory \cite{Werthamer1966}. 

The spin paramagnetic pair breaking mechanism is related to the Zeeman splitting of the spin 
singlet Cooper pairs due to the interaction of the magnetic field with electron spins. 
For a BCS superconductor, the Pauli limiting field is given by
 $H{^\mathrm{Pauli}} = 1.82T{_c} = 5.80T$. 
The upper critical field $\mu{_0}H{_{\mathrm{c2}}}$ is influenced by both orbital and spin paramagnetic 
effects.  Orbital pair breaking is the dominant mechanism at low magnetic fields and Pauli 
paramagnetic effect dominates the upper critical field at very high magnetic fields. The relative 
importance of the Orbital and Pauli limiting fields is described by the Maki parameter 
`$\alpha$' \cite{Maki1966} defined as, 
\begin{equation}\label{eq:4}
\alpha=\sqrt{2}{H{_{c2}}^{\mathrm{orb}}(0)}/H{^\mathrm{Pauli}}(0).
\end{equation}

The value of the orbital critical field calculated using equation~(\ref{eq:2}) is 5.45~T. 
The Ginzburg-Landau coherence length, $\xi(0){_{\mathrm{GL}}}$ is given by,
\begin{equation}\label{eq:5}
\xi(0){_{\mathrm{GL}}}=\sqrt{\Phi{_0}/{2\pi{H{_{c2}}^{\mathrm{orb}}(0)}}}.
\end{equation}

Using equation~(\ref{eq:5}), we get $\xi(0){_{\mathrm{GL}}} = 78\AA$. Using the values of 
$\mu{_0}{H{_{\mathrm{c2}}}^{\mathrm{orb}}(0)}$ and $\mu{_0}H{^\mathrm{Pauli}}$,
the value of the Maki parameter comes out to be $\alpha=1.33$.

The temperature dependence of $H_{\mathrm{c2}}$ for single-band, dirty limit superconductors is given by 
the WHH formula,
\begin{equation}\label{eq:6}
ln\left(\frac{1}{t}\right)=\sum_{\nu=-\infty}^{\nu=\infty}\Biggl\{\frac{1}{|2\nu+1|}-\Biggl[{|2\nu+1|}+
\frac{\overline h}{t}+
\frac{(\alpha\overline h/t)^2}{|2\nu+1|+(\lambda_{so}+\overline h)/{t}}\Biggr]^{-1}\Biggr\},
\end{equation}
where $t=T/T{_c}$, $\overline h = {4H_{c_2}(T)}/({\pi^2T_c \big| \frac{dH_{c_2}(T)}{dT}\big|_{T_c}})$, 
$\alpha$ is the Maki parameter, and $\lambda_{so}$ is the spin-orbit scattering constant. When 
$\lambda_{so}=0$, ${H{_{c2}}(0)}$ obtained from WHH formula satisfies the relation,
\begin{equation}\label{eq:7}
{H{_{\mathrm{c2}}}(0)}={H{_{c2}}^{\mathrm{orb}}(0)}/{\sqrt{1+\alpha^2}}.
\end{equation}

In Fig.~\ref{fig:fig2}(b), we note that the experimental ${H{_{c2}}}$ vs T data is best described 
by the WHH formula for $\alpha=0$ ($H{_{c2}}^{\mathrm{orb}}(0) \ll H{^\mathrm{Pauli}}(0)$),   $\lambda_{so} = 0$
(red dashed curve). If we use $\alpha=1.33$, as derived 
above, the WHH formula does not explain the data except when $\lambda_{so}\approx{100}$, which is clearly unphysical. This suggests that single band models are not enough to describe the temperature dependence of upper critical field ${H{_{c2}}}(T)$ and multi-band effects have to be included in the models to describe the data. The multi-band models \cite{Gurevich2010} need parameters like inter-band coupling constants and knowledge of Matsubara frequencies for the specific compound and these are not yet known for Lu$_3$Os$_4$Ge$_{13}$. Assuming moderate multi-band effects present in this compound, if we do a linear extrapolation of the ${H{_{c2}}}(T)$ at T=0 in Fig.~\ref{fig:fig2}(b), we get an estimate of the ${H{_{c2}}}(0)=6.8T$, which is $\approx$ 1T more than the value obtained from WHH theory. With this observation, to an first order approximation, we can use single band models to get approximate values of superconducting state parameters, like coherence length, penetration depth etc. 
\begin{figure}[h]
\includegraphics[width=16cm,angle=0]{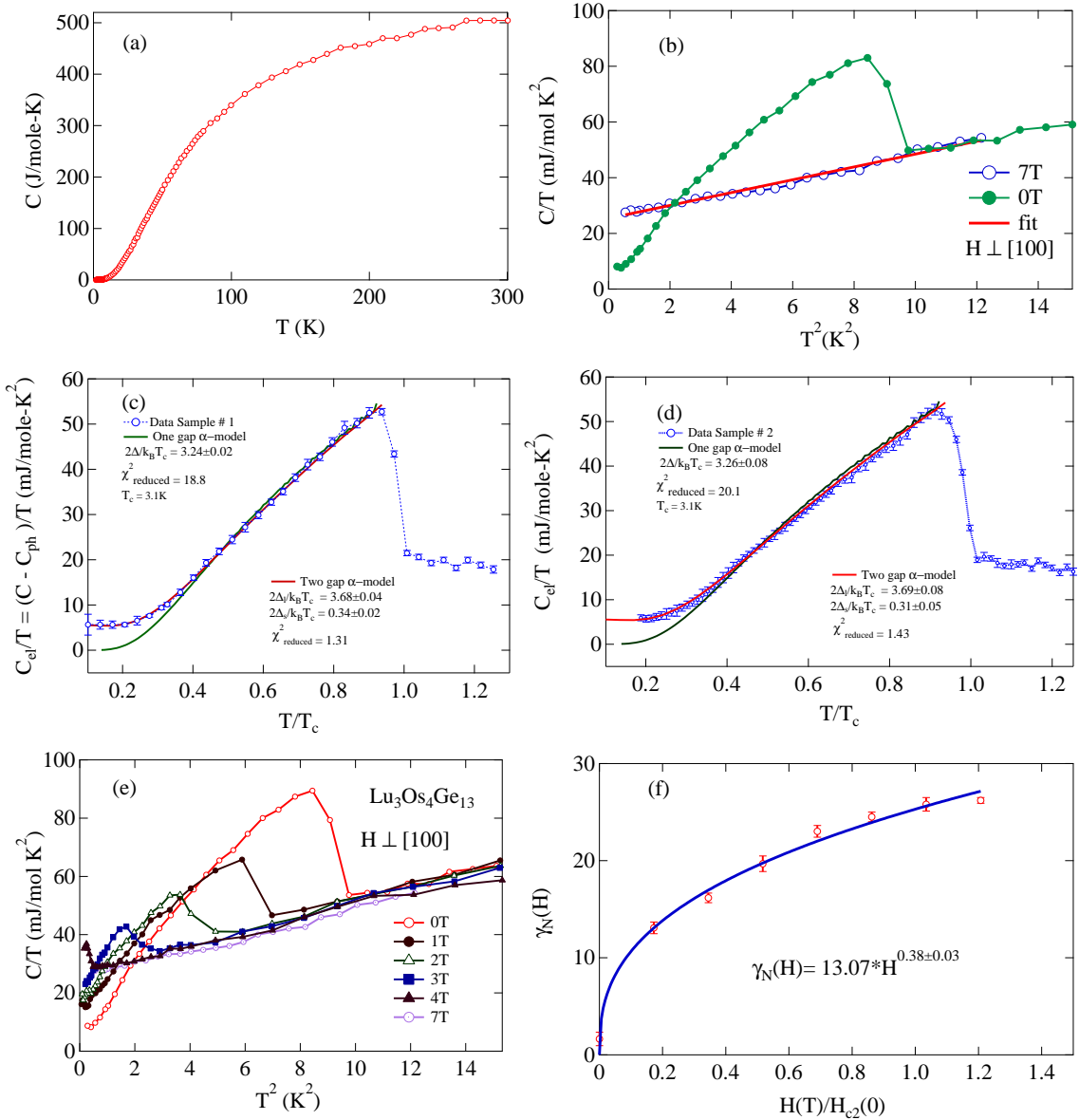}
\caption{(Color online) (a) Specific heat capacity ($C$) vs temperature of Lu$_3$Os$_4$Ge$_{13}$ from 
1.8-300~K.
(b) Normal state heat capacity data taken at 7T fitted to the equation $C(T) = \gamma{_N} T+ \beta T^3$. The zero field data shows sharp jump at 3.1K. 
(c) Low temperature electronic heat capacity data for sample $\#$1 fitted using: (i) one-gap $\alpha$-model (anisotropic gap) and 
(ii) Two-gap $\alpha$-model for a multi-band BCS superconductor. 
(d) Low temperature electronic heat capacity data for sample $\#$2 fitted using: (i) one-gap $\alpha$-model (anisotropic gap) and 
(ii) Two-gap $\alpha$-model for a multi-band BCS superconductor. 
(e) The heat capacity in the mixed state in different magnetic fields. The linear extrapolations at T=0 of the low temperature heat capacity data at different fields are used to estimate the values of Sommerfeld coefficient $\gamma {_N} (H)$ in the superconducting state.
(f) The magnetic field dependence of $\gamma {_N} (H)$.}

\label{fig:fig3}
\end{figure}

Fig.~\ref{fig:fig3}(a) shows the zero field heat capacity data of Lu$_3$Os$_4$Ge$_{13}$ in the 
temperature range 1.8-300K. 
The experimental value of heat capacity 
$C(T=300K)=504$ Jmole${^{-1}}$K${^{-1}}$ is very close to the Dulong-Petit high-temperature limit 
of the lattice heat capacity $C{_v} = 3N{\it{R}} = 498.9$~Jmole${^{-1}}$K${^{-1}}$.
The electronic contribution to the heat capacity $C{_{el}}(T)$ can be calculated by subtracting 
the phonon contribution from the total heat capacity $C(T)$, i.e. $C{_{el}}(T) = C(T)-{\beta T^3}$. 
Fig.~\ref{fig:fig3}(b) shows a sharp jump in the heat capacity at 3.1~K which confirms the bulk 
superconductivity in the compound. Fig.~\ref{fig:fig3}(e) shows the heat capacity data measured in 
different magnetic fields ($H \perp [100]$). The magnitudes of the heat capacity jump 
as well as $T{_c}$ decrease with increasing magnetic field. 

The superconductivity is fully suppressed in a magnetic field of $7~T$ and the data is fitted to 
the equation $C(T) = \gamma{_n} T + \beta T^3$ as shown in Fig.~\ref{fig:fig3}(b), where 
$\gamma{_n} T$ represents the electronic contribution and $\beta T^3$ describe the lattice-phonon 
contribution to the specific heat in the normal state. We find the electronic 
specific heat coefficient $\gamma{_n} = 25.4\pm {0.3} \frac{mJ}{mol K^2}$ and the phonon/lattice contribution 
coefficient $\beta = 2.30 \pm {0.05}\frac{mJ}{mol K^4}$ from the fit. The Debye temperature ($\Theta{_{\mathrm{D}}}$), 
calculated using the formula,
\begin{equation}\label{eq:8}
\Theta{_{\mathrm{D}}} = ({12\pi{^4}{\it{R}}N}/{5\beta}){^{\frac{1}{3}}},
\end{equation}
where {\it{R}} is the molar gas constant and N($=20$) is the number of atoms per formula unit (f.u.), 
is $256.6\pm0.7$~K.
The density of states at the Fermi level calculated using the formula $\gamma{_n}=(\pi{^2}{k{_B}}{^2}/3)D(E{_F})$, 
is $D(E{_F}) = 21.3$ states/eV-f.u. for both spin directions. This density of state contains 
quasiparticle mass enhancement by many-body electron-phonon interaction and is related to bare 
density of states $D{_{band}}(E{_F})$ by $D(E{_F})=(1+\lambda{_{\mathrm{eph}}})D{_{\mathrm{band}}}(E{_F})$, 
where $\lambda{_{\mathrm{eph}}}$ is the dimensionless 
electron-phonon coupling constant. $\lambda{_{\mathrm{eph}}}$ is related to the phonon spectrum and density
of states in Eliashberg theory \cite{Eliashberg1960} and represents the strength of electron-phonon coupling. 
$\lambda{_{\mathrm{eph}}}$ can be calculated using McMillan's formula \cite{McMillan1968},
\begin{equation}\label{eq:9}
\lambda{_{\mathrm{eph}}}=\frac{1.04+\mu^*ln(\Theta{_\mathrm{D}}/1.45T{_c})}{(1-0.62\mu^*)ln(\Theta{_{\mathrm{D}}}/1.45T{_c})-1.04},
\end{equation}
where $\mu^*$ is the repulsive screened coulomb parameter. The competition between $\mu^*$ and 
$\lambda{_{\mathrm{eph}}}$ is the determining factor for Cooper pairing in conventional superconductors 
\cite{Bardeen1957}. The value of $\mu^*$ is taken as 0.13. Using equation~(\ref{eq:9}), we obtain 
$\lambda{_{\mathrm{eph}}}$ = 0.58. This suggests a moderately enhanced electron-phonon coupling in Lu$_3$Os$_4$Ge$_{13}$. 
Combining the values of $\lambda{_{\mathrm{eph}}}$ and $D(E{_F})$, we get 
$D{_{band}}(E{_F})=13.48$ states/eV-f.u. for both spin directions.
The effective mass of the quasiparticles ($m^*$), calculated using the relation, 
$m^*=(1+\lambda{_{\mathrm{eph}}})m{_{\mathrm{band}}}$ is $m^* = 1.58m{_e}$,
assuming effective band mass $m{_{\mathrm{band}}}=m{_e}$, the free electron mass.

The sharp jump in the electronic heat capacity $\Delta C{_{el}}(T)$ at $T{_c}$, is 89.6~mJ/mol K, 
giving the ratio ${\Delta C{_{\mathrm{el}}}}/{\gamma{_n} T{_c}}= 1.15$. 
This ratio can be used to understand the strength of
the electron-phonon coupling. This ratio is smaller 
than the weak-coupling limit value of 1.43 for a conventional BCS superconductor. 
If we consider a single band model \cite{Padamsee1973} for a BCS superconductor, 
such a reduction in the heat capacity jump can be either due to the presence of 
anisotropic superconducting energy gap or presence of multiple gaps at the Fermi 
surface in the superconducting state. 

Apart from the reduced jump in the heat capacity at $T{_c}$, the low temperature electronic heat 
capacity is higher than the expected value for a single gap s-wave BCS superconductor 
(see Fig.~\ref{fig:fig3}(c) and (d)). To analyze the suppression in the heat capacity jump and the
low temperature data in more detail, we have used empirical (and not self-consistent) one-band one-gap 
${\alpha}$-model (which accounts for anisotropy in the order parameter $\Delta{_0}$ at 
the Fermi surface) and two-band two-gap $\alpha$-model \cite{Padamsee1973, Kogan2009}. In these models, the superconducting 
energy gap ($\Delta(t)$) is parametrised in terms of normalised BCS gap ($\delta(t)$), $\Delta(t) =  \Delta{_0} \delta(t)$, 
where t ($=T/T{_c}$) is the reduced temperature.
The normalised BCS gap as a function of reduced temperature is taken from the Muhlschlegel's paper \cite{{Muhlschlegel}}. The entropy (S) and the heat capacity (C) 
for a system of independent fermionic-quasiparticles can be written as,
\begin{equation}\label{eq:10}
\frac{S}{\gamma{_n} T{_c}} = -\frac{6}{{\pi}^2} \frac{\Delta{_0}}{k{_B}T{_c}} \int_0^{\infty} [f\ln f + (1-f)\ln (1-f)]dy,
\end{equation}
\begin{equation}\label{eq:11}
\frac{C_{el}}{\gamma{_n} T{_c}} = t \frac {d({S}/{\gamma{_n} T{_c}})}{dt},
\end{equation}
where $f = [e^{\beta E}+1]^{-1}$, $\beta = (k{_B}T)^{-1}$ and $y = \varepsilon /\Delta{_0}$. The quasiparticle energy (E) is 
$\sqrt{{\varepsilon }^2 + {\Delta}^2 (t)}$, where $\varepsilon $ is the energy of the 
normal electrons measured from the Fermi level. Using equation~(\ref{eq:10}) and (\ref{eq:11}), we can write the  
electronic heat capacity for one-gap $\alpha$-model as,
\begin{equation}\label{eq:12}
C_{el}(a,\alpha, t)=a\int_0^{\infty}\Biggl[\bigg (\frac{x}{t}\bigg)^2+{\alpha}^2 \bigg(\frac{\delta(t)}{t}\bigg)^2-{\alpha}^2\bigg(\frac{\delta(t)}{t}\bigg)\bigg(\frac{d\delta(t)}{dt}\bigg)\Biggl]\\
{\sech}^2\left(\sqrt{\left(\frac{x}{t}\right)^2+{\alpha}^2\left(\frac{\delta(t)}{t}\right)^2}\right){dx},
\end{equation}
where $a= {12 \gamma{_{n}} T{_c}}/{{\pi}^2}$, $x = \varepsilon /{2 k{_B}T{_c}}$ and $\alpha = \Delta / {2 k{_B}T{_c}}$.
In the two-gap model, the total electronic heat capacity, $C_{tot}(a{_{1}},\alpha{_1},a{_{2}},\alpha{_2},t)$, is taken as the sum of the independent contributions 
from two-bands with different superconducting energy gaps, each following BCS type 
temperature dependence (if we neglect inter-band 
transitions due to scattering by impurities or phonons, and assume that $\gamma{_{n,l}} + 
\gamma{_{n,s}} = \gamma{_n}$). For the two-gap $\alpha$-model, the total heat capacity is given by,
\begin{equation}\label{eq:13}
C_{tot}(a{_{1}},\alpha{_1},a{_{2}},\alpha{_2},t)=C_{el}(a{_{1}},\alpha{_1},t)+C_{el}(a{_2},\alpha{_2},t),
\end{equation}
where $a{_1} = {12 \gamma{_{n,l}} T{_c}}/{{\pi}^2}$, $\alpha{_1} = \Delta{_{l}} / {2 k{_B}T{_c}}$,
$a{_2} = {12 \gamma{_{s,l}} T{_c}}/{{\pi}^2}$ and $\alpha{_2} = \Delta{_{s}} / {2 k{_B}T{_c}}$. 
The subscripts `l' and `s' stand for large and smaller gap respectively.

We estimate the fitting parameters ($a$, $\alpha$ for the one-gap $\alpha$-model; and $a{_{1}}$, $\alpha{_1}$, $a{_{2}}$, $\alpha{_2}$ for two-gap $\alpha$-model) using nonlinear 
least-squares fit for two different samples in order to confirm the reproducibility of the results. To estimate the uncertainties in the fitted parameters, we repeat the whole process for 
1000 realizations of the data obtained by adding random perturbations to the data with the standard 
deviation equal to $1\sigma$ uncertainty in the corresponding data point for both samples. The median value of each 
parameter for 1000 realizations was accepted as the fitted value, while the $\pm 1\sigma$ uncertainty 
in the fitted value was estimated from the range covering 34$\%$ of the area on either side of
the median in the distribution function of the fitted parameter. Fig.~\ref{fig:fig3}(c) and fig.~\ref{fig:fig3}(d) show
the fit to the electronic specific heat for the two samples 
using one-gap $\alpha$-model and two-gap $\alpha$-model. The one-gap $\alpha$-model fails to fit the data at 
the low temperatures with large values of chi-square per degree of freedom ($\chi^2_{\mathrm{pdf}} = 18.8$ and 20.1 for sample $\#$1 and sample $\#2$ respectively). The low temperature electronic heat capacity of both samples is best described by the two-gap $\alpha$-model with $\chi^2 _{\mathrm{pdf}}= 1.31 (1.43)$. Clearly, the two-gap $\alpha$-model fits the data significantly better than 
the one-gap $\alpha$-model.  We find the two gaps to be $2\Delta{_{l}} / {k{_B}T{_c}} = 3.68 \pm {0.04} ~(3.69 \pm {0.08})$ and 
$2\Delta{_{s}} / {k{_B}T{_c}} = 0.34 \pm {0.02}~ (0.31 \pm {0.05})$ for sample $\#$ 1 (sample $\#$ 2). The contributions of $\gamma{_{n,l}}$ and $\gamma{_{n,s}}$ 
to $\gamma{_n}$ are $81.5\%$ and $18.5\%$ respectively. 

The study of vortex excitations in the mixed state provides insight in the understanding of the 
superconducting order parameter. Fig.~\ref{fig:fig3}(e) shows the low temperature electronic heat 
capacity in different magnetic fields ($H_{c1} < H < H_{c2}$). The linear extrapolations to zero temperature 
give the heat capacity contribution due to normal electrons in the mixed state, and are determined 
in terms of the electronic specific heat coefficient $\gamma{_N}(H)$. This electronic 
contribution is attributed to the normal state electrons present in the vortex 
cores. In s-wave superconductors, the vortex cores contribute to the electronic 
heat capacity as normal metals. This contribution is proportional to the number of vortices and 
hence proportional to the applied magnetic field, i.e. $\gamma{_N}(H) \propto H$. However, 
we find that $\gamma{_N}(H) = 13.07H^{0.38\pm0.03}$ as shown in the Fig.~\ref{fig:fig3}(f). 
A nonlinear dependence of $\gamma{_N}(H)$ on the magnetic field has been argued to be intrinsic property of the 
multi-band, multi-gap superconductors \cite{yang}. This analysis also suggests the presence of multiple gaps 
at the Fermi level in the superconducting state and indicate multi-band superconductivity in 
Lu$_3$Os$_4$Ge$_{13}$ single crystal. 
\begin{figure}[h]
\includegraphics[width=16cm,angle=0]{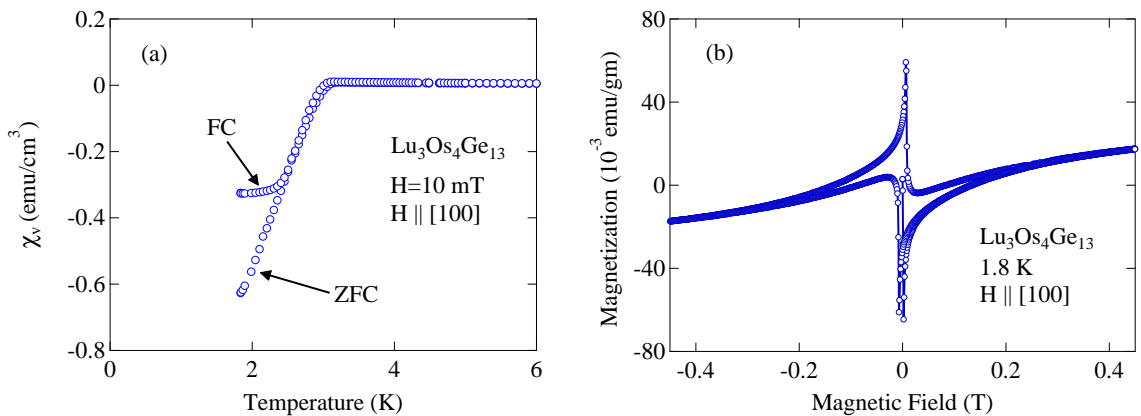}
\caption{(Color online) (a) DC magnetic susceptibility data as function of temperature. The superconducting 
transition temperature, as determined from susceptibility measurement is in excellent agreement with 
the resistivity data. The ZFC and FC susceptibility data indicate significant amount of pinning of 
vortices in the compound. 
(b) Magnetization as a function of magnetic field for H $\parallel$ [100] direction. The M-H loop is closed 
(${\delta M=0}$) at magnetic fields $H\ge{0.45~T}$ at 1.8~K, which suggests possible melting of vortices at 0.45~T field.}
\label{fig:fig4}
\end{figure}

Fig.~\ref{fig:fig4}(a) shows a diamagnetic transition into superconducting state 
at 3.1~K in the low temperature dc-susceptibility data. Significant amount of vortex pinning can be observed 
by comparing the zero field cooled (ZFC) and the field cooled (FC-Meissner) data below the 
transition temperature. The sample is weakly paramagnetic at room temperature 
($\chi = 3\times10^{-4}$ emu/mole at T=300K). The temperature dependence of the susceptibility is 
Pauli paramagnetic type with small rise at low temperature ($\chi = 8\times10^{-3}$ emu/mole at T=3.2K) 
possibly due to the presence of small magnetic rare-earth impurities (ppm level) in Lu. We estimate the value of 
the lower critical field $H{_{c1}}(1.8K)=20$~mT from magnetization measurements. The 
value of the lower critical field $H{_{c1}}(0)=30$~mT is obtained using the following expression,
\begin{equation}\label{eq:14}
H{_{c1}}(0)=H{_{c1}}(T)/{(1-(T/T_{c})^2)}.
\end{equation}
Substituting the values of $\xi_{GL}(0)$ and $H{_{c1}}(0)$ in the following expression,
\begin{equation}\label{eq:15}
H_{c1}(0)=\frac{\Phi_{0}}{{4 \pi {{\lambda}^2_{GL}}}}\ln(\frac{\lambda_{GL}}{\xi_{GL}}),
\end{equation}
we obtain $\lambda_{GL}(0)=4736~\AA$. The Ginzburg-Landau parameter 
$\kappa_{GL}(0)=\frac{\lambda_{GL}(0)}{\xi_{GL}(0)}=61$. Using the formula, 
\begin{equation}\label{eq:16}
H{_{c1}}(0)=\frac{H_{c}(0)}{\sqrt{2}\kappa}(\ln{\kappa}+0.5),
\end{equation}
we obtain the value of the thermodynamic critical field $H{_{c}}(0) = 564~mT$. The magnetization 
(M-H) data as shown in Fig.~\ref{fig:fig4}(b) shows that the M-H loop is closing at magnetic 
fields H~${>0.45}$~T at 1.8~K, suggesting the unpinning (melting) of vortices in magnetic 
fields much smaller than the of upper critical field ($~\mu{_0}H{_{c2}}(1.8K)=3.2~T$). 
An unpinned vortex phase exists in the magnetic field region 
$0.45~T\le H \le {\mu{_0}H{_{c2}}(1.8~K)=3.2~T}$, which may be a vortex liquid phase. 
Similar vortex liquid phase is observed in high $T{_c}$ superconductors \cite{Blatter1994} 
and is an unusual phenomena in low $T{_c}$ superconductors.
Though the melting of the vortices in high $T{_c}$ 
superconductors is attributed to quantum and thermal fluctuations \cite{Baruch2010,Blatter1994}, 
there is no clear understanding of the origin of vortex liquid phase in low $T{_c}$ superconductors. 
The strength of the thermal fluctuations, which leads to the vortex unpinning, is described in 
terms of the Ginzburg number given by,
\begin{equation}\label{eq:17}
G_{i}=\frac{1}{2}\left(\frac{k_{B}\mu_{0}\Gamma T_{c}}{4\pi\xi^3(0)H^2_{c}(0)}\right)^2,
\end{equation}
where $\Gamma$ is the anisotropy parameter ($\approx{1}$ for cubic Lu$_3$Os$_4$Ge$_{13}$). 
Using equation~(\ref{eq:17}), we get the value of the Ginzburg number $G_{i}=4.1\times10^{-6}$. 
The value of Ginzburg number 
for Lu$_3$Os$_4$Ge$_{13}$ is larger than low $T_{c}$ superconductors ($\approx{10^{-8}}$) but 
smaller than high $T_{c}$ superconductors ($\approx{10^{-2}}$) \cite{Baruch2010,Blatter1994}. 
This suggests that the thermal fluctuations, though weak, may play an important role in the unpinning of the 
vortices in this compound.

Type-II superconductors in which $H_{c1}$ and $H_{c2}$ are well separated ($H_{c2}/H{_{c1}}\sim\kappa^{2}$) 
are classified as strongly type-II superconductors. This ($H_{c1}\ll {H_{c2}}$) leads to a situation in which 
the magnetic fields associated with vortices overlap and the superposition becomes nearly homogeneous, 
while the order parameter characterizing superconductivity is still inhomogeneous \cite{Baruch2010}. 
Detailed penetration depth ($\lambda_{GL}$) measurements are required to obtain correct values of 
$H_{c1}$ and $\kappa$.

The analysis of the electronic band structure and density of states of Lu$_3$Os$_4$Ge$_{13}$, based 
on the electronic structure calculations using Wien2K, is presented below. The left panel in Fig.~\ref{fig:fig5} shows the 
band structure of Lu$_3$Os$_4$Ge$_{13}$ near the Fermi level. The band structure has multi-valley type character 
\cite{Cohen1964}. Three bands cross the Fermi level and account for metallic nature of the compound. 
The right panel in Fig.~\ref{fig:fig5} shows the calculated density of states (DOS) for one formula unit (20 atoms) 
of Lu$_3$Os$_4$Ge$_{13}$. 
\begin{figure}[h]
\includegraphics[width=15cm,angle=0]{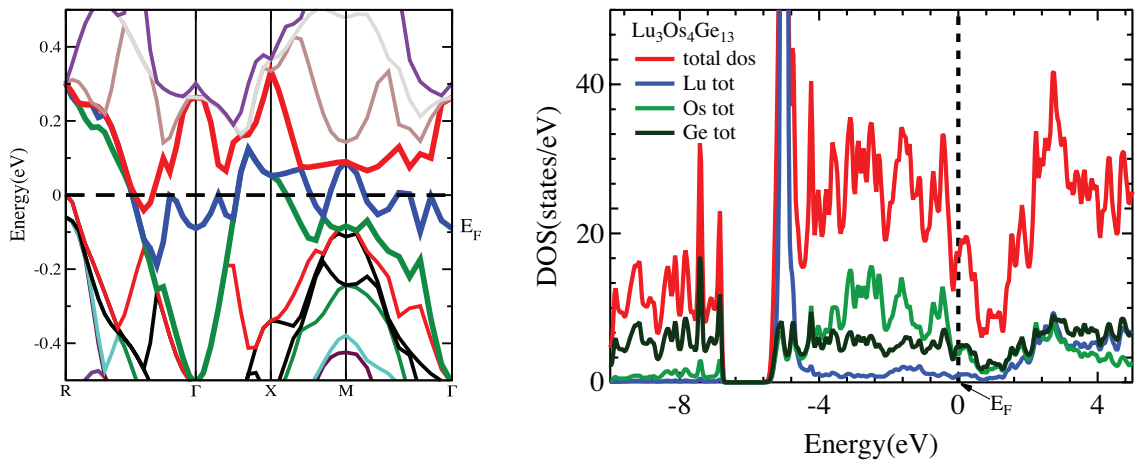}
\caption{(Color online) ({\it Left panel}) Band structure of Lu$_3$Os$_4$Ge$_{13}$ near Fermi level. The 
band structure shows a multi-valley type character. Three bands shown in bold lines cross the Fermi-surface. 
({\it Right panel}) Analysis of density of states of Lu$_3$Os$_4$Ge$_{13}$. The total DOS curve has a local maximum at the edge of 
the Fermi energy. The partial density of states curves show that the major contribution to the total 
density of states is coming from Os and Ge atoms. Lu atoms have least contribution to the total density of states.}
\label{fig:fig5}
\end{figure}
\begin{figure}[h]
\includegraphics[width=15cm,angle=0]{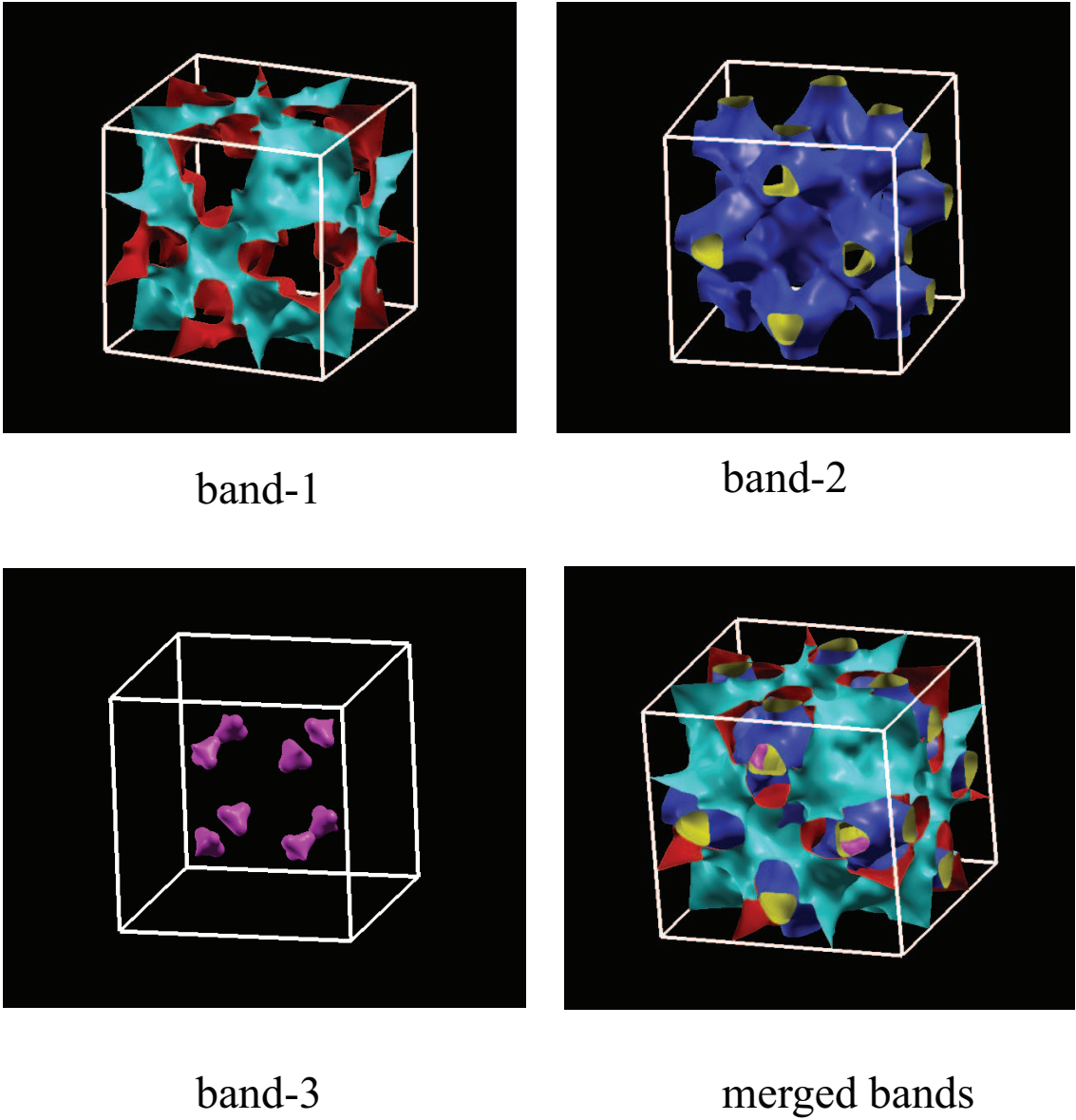}
\caption{(Color online) Contributions of the bands crossing the Fermi level to the Fermi surface. 
All the three bands are plotted together make a complex surface.}
\label{fig:fig6}
\end{figure}

The Fermi level is located near the edge of a local maxima in total density of states. 
The value of the DOS at $E{_F}$ is $\approx{17}$ states/eV-f.u for both spin directions. 
This value is smaller than the value ($\approx{21}$ states/eV-f.u for both spin directions) calculated using the value of
${\gamma{_n}}$ obtained from heat capacity measurements, which indicates the mass-enhancement in the compound, 
since no electronic correlations were taken into account in the band structure calculations. The partial 
DOS shows that the total DOS is dominated by contributions from Os and Ge. Fig.~(\ref{fig:fig6}) shows 
the calculated Fermi surfaces for the three bands (as well as all the bands plotted together) 
which cross the Fermi level. The combination of these bands leads to a 
complex Fermi surface (Fig.~\ref{fig:fig6}). 
\section{Conclusion}
In summary, we have studied the superconducting properties of Lu$_3$Os$_4$Ge$_{13}$ in detail using electrical transport, magnetization and heat capacity measurements. We show that Lu$_3$Os$_4$Ge$_{13}$ is multi-band type-$II$ superconductor ($T_{c}=3.1$K) by analyzing the low temperature heat capacity data using empirical two-gap multi-band $\alpha$-model. The single band WHH model does not fully explain the temperature dependence of the upper critical field $H{_{\mathrm{c2}}}(T)$, suggesting presence of multi-band effects in Lu$_3$Os$_4$Ge$_{13}$. The analysis of the low temperature heat capacity data for two different samples of Lu$_3$Os$_4$Ge$_{13}$ single crystal using two-gap $\alpha$-model confirms the presence of two superconducting gaps ($2\Delta{_{l}} / {k{_B}T{_c}} = 3.68 \pm {0.04} (3.69 \pm {0.08})$ and $2\Delta{_{s}} / {k{_B}T{_c}} = 0.34 \pm {0.02} (0.31 \pm {0.05})$) in the compound. The magnetization measurements show a large reversible region in the mixed state, similar to the vortex liquid phase observed in high-$T_{c}$  superconductors. The estimation of the Ginzburg number $G_{i}$ suggests that thermal fluctuations (though small) may play an important role in the unpinning of the vortices in this compound. Electronic band structure calculations along with heat capacity measurements suggest that the electronic correlations are not significant in Lu$_3$Os$_4$Ge$_{13}$. Band structure calculations show a very complex structure in the fermi surface which might play significant role in enhancing multi-band effects in Lu$_3$Os$_4$Ge$_{13}$.
\section{Acknowledgments}

We thank Dr. Susanta K. Mohanta, Prof. Surendra Nath Mishra, Dept. of Nuclear $\&$ Atomic Physics, 
Dr. Sutirtha Mukhopadhyay, Prof. Pratap Raychaudhuri, Dept. of Condensed Matter Physics $\&$ Material 
Science and Kuldeep Verma, Dept. of Astronomy and Astrophysics, TIFR for useful discussions.

\end{document}